\documentclass[a4paper,11pt]{article}
\usepackage{pos}
\usepackage{setspace}
\usepackage{lineno}

\title{$\psi(2S)$ production in Pb-Pb collisions measured by ALICE at the LHC~\cite{ALICE:2022jeh}}
\ShortTitle{$\psi(2S)$ production in Pb-Pb collisions}

\author*[a]{Victor Feuillard}
 \onbehalf{on behalf of the ALICE collaboration}

\affiliation[a]{Physikalisches Institut, \\
  Im Neuenheimer Feld 226, 69120 Heidelberg, Germany}

\emailAdd{victor.jose.gaston.feuillard@cern.ch}

\abstract{Charmonium production is a probe sensitive to deconfinement in nucleus-nucleus collisions. The production of J/$\psi$ via regeneration within the QGP or at the phase boundary has been identified as an important ingredient for the description of the observed J/$\psi$ nuclear modification factor at the LHC.  The $\psi(2S)$ production relative to J/$\psi$ is a possible discriminator between the two regeneration scenarios. Studies of $\psi(2S)$ production in central nucleus-nucleus collisions at low transverse momentum ($p_{\rm T}$) are crucial, particularly at the LHC, where regeneration appears to be dominant. A significant $\psi$(2S) signal is extracted at low $p_{\rm T}$ and forward rapidity in the dimuon decay channel for the first time. This measurement relies on the $\psi(2S)$ cross section measured recently in pp collisions at  $\sqrt{s_{\rm }}=5.02$~TeV with an unprecedented precision compared to previous ALICE results. In this contribution, we present newly published results on the $\psi(2S)$-to-J/$\psi$ ratio and the $\psi(2S)$ nuclear modification factor in Pb--Pb collisions at $\sqrt{s_{\rm NN}}=5.02$~TeV. Results are reported as a function of centrality and $p_{\rm T}$ for $p_{\rm T}<12$~GeV/c and are compared to available NA50 and CMS measurements. Comparisons to transport and statistical hadronization model predictions are also provided to shed light on the charmonium states recombination mechanism.}

\FullConference{HardProbes2023\\
 26-31 March 2023\\
 Aschaffenburg, Germany\\}

\begin{document}
\maketitle
The quark--gluon plasma (QGP) is a state of matter theoretically predicted by quantum chromodynamics (QCD) where quarks and gluons are deconfined and chiral symmetry is restored at extremely high temperature ($T_c \approx 155$~MeV with zero net baryonic number~\cite{Gross:2022hyw}). There is a particular interest in studying the QGP since, according to the current cosmological models, our Universe was in such a deconfined state in its early stages, after the electroweak phase transition and up to $\tau\approx10$~$\mu$s after the Big Bang. Experimentally, it is possible to create the QGP through ultra-relativistic heavy-ion collisions, such as those happening at the SPS~\cite{Angert:249000}, RHIC~\cite{ROSER200223}, or the LHC~\cite{LyndonEvans_2008}, but for only a short period of time and within a very small volume (e.g. $\tau\sim$10~fm/\textit{c} and $V\sim5\cdot10\textsuperscript{3}$~fm\textsuperscript{3} in Pb--Pb collisions at $\sqrt s\textsubscript{NN} = 2.76$~TeV)~\cite{ALICE:2011dyt}. 

Charmonium resonances such as J/$\psi$ and $\psi(2S)$ are bound states of a $\rm{c\bar{c}}$ pair. Because of their large mass, charm quarks are primarily produced in hard scatterings at the beginning of the collision. Therefore, they experience the entire medium evolution and are affected by it, which makes charmonia among the most direct signatures to investigate QGP properties. Charmonia can be affected by the QGP in several ways. Firstly, theory predicts that quarkonia are suppressed in a QGP due to color screening~\cite{MATSUI1986416} and dynamical dissociation~\cite{Rothkopf:2019ipj}. This leads to a reduced number of charmonium states produced in Pb--Pb collisions with respect to scaled pp collisions. Secondly, a competing mechanism can occur, namely (re)combination: if there are enough heavy-quark pairs produced, then quarkonia can be regenerated by the combination of these quarks either at the phase boundary~\cite{Braun-Munzinger:2000csl} or during the QGP phase~\cite{Thews:2000rj}, increasing the charmonium yields. It is particularly interesting to compare the J/$\psi$ and $\psi(2S)$ production, because although they are both charmonium state, the binding energy of the $\psi(2S)$ is ten times lower than the binding energy of the J/$\psi$~\cite{Satz:2005hx}. Therefore, it is expected that the $\psi(2S)$ will be dissociated at a lower medium temperature than the J/$\psi$. Concomitantly, the J/$\psi$ and the $\psi(2S)$ differ in size by a factor of 2. This could lead to different recombination processes, the $\psi(2S)$ being produced in some models at later stages of the QGP evolution, when the system is more dilute. Therefore, studying both J/$\psi$ and $\psi(2S)$ can provide insightful information to test the recombination models.

\begin{figure}[!htb]
\begin{center}
\vspace{9pt}
\includegraphics[scale=0.4]{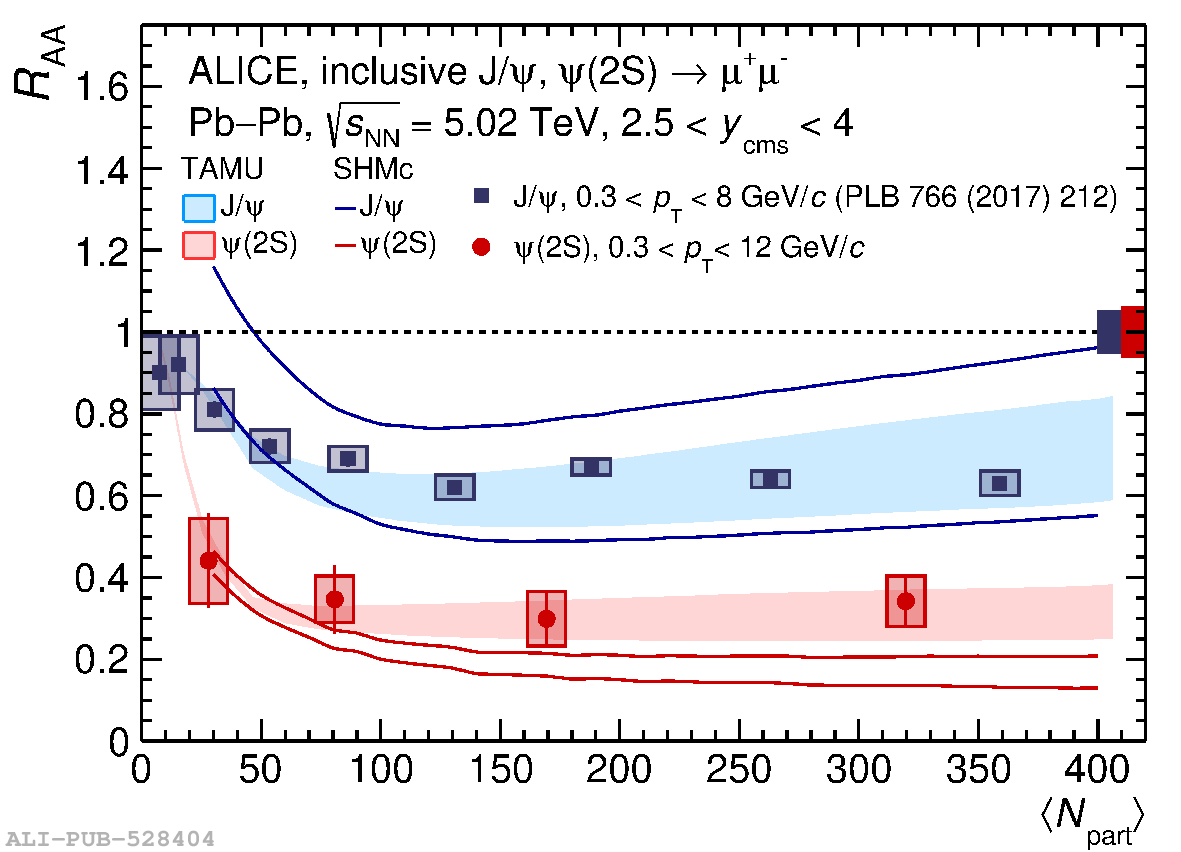}
\caption{Nuclear modification factor of $\psi(2S)$~\cite{ALICE:2022jeh} and J/$\psi$~\cite{ALICE:2016flj} as a function of centrality. The values are compared with model predictions from SHMc~\cite{Andronic:2019wva} and TAMU~\cite{Du:2015wha}.}
\label{fig:Figure1}
\end{center}
\end{figure}

The results presented here are obtained using the muon spectrometer of ALICE, which allows one to measure inclusive charmonium in the dimuon decay channel at forward rapidity ($2.5<y<4.0$), down to $p_{\rm T}=0$. A complete description of the ALICE detector can be found in Ref~\cite{ALICE:2008ngc,ALICE:2014sbx}. The data was collected during 2 Pb--Pb data taking periods in 2015 and 2018, corresponding to a total integrated luminosity of $L_{\rm int} = 750$~$\rm{\mu b^{-1}}$. The nuclear modification factor $R_{\rm AA}$ is defined as $R_{\rm AA} = \frac{\rm Y_{PbPb}}{\langle T_{\rm AA}\rangle \cdot \rm{\sigma_{pp}}}$, where $\rm{Y_{PbPb}}$ is the charmonium yield in Pb--Pb collisions, $\langle T_{\rm AA}\rangle$ is the nuclear overlap function that is linked to the centrality of the collision, and $\sigma_{\rm pp}$ is the $\psi(2S)$ production cross-section in pp collisions in the same acceptance and at the same center-of-mass energy, reported in Ref~\cite{ALICE:2021qlw}. The charmonium yield is extracted employing a $\chi^2$ minimization fit to the opposite sign dimuon invariant mass spectrum, in which the combinatorial background is subtracted with the help of an event mixing procedure and the resonance signals are described by a double-sided Crystal Ball function or a pseudo-Gaussian with a mass-dependent width. The $\psi(2S)$ $R_{\rm AA}$ as a function of centrality is presented in Figure~\ref{fig:Figure1}, compared with the J/$\psi$ $R_{\rm AA}$~\cite{ALICE:2016flj}. A $p_{\rm T}>0.3$~GeV/$c$ cut is applied to remove the contribution from photoproduction in peripheral collisions.

Whereas J/$\psi$ exhibits a smaller suppression in peripheral collisions and a constant $R_{\rm AA}$ in mid-central and central collisions, the $\psi(2S)$ $R_{\rm AA}$ shows no centrality dependence within the uncertainties. Moreover, a larger suppression of the $\psi(2S)$ with respect to the J/$\psi$ is observed in the entire centrality range, which is consistent with the expectations. The $R_{\rm AA}$ is also compared with theoretical predictions. First, the Statistical Hadronization Model (SHMc)~\cite{Andronic:2019wva} assumes no binding of charmonia in the QGP phase and that the charmonium production occurs at the phase boundary by the statistical hadronization of charm quarks. On the other hand, the Transport Model (TAMU)~\cite{Du:2015wha} assumes continuous charmonium dissociation and regeneration in the QGP, described by a rate equation. The TAMU prediction shows a good agreement with the $R_{\rm AA}$ for both J/$\psi$ and $\psi(2S)$ within the uncertainties. The SMHc model is also able to reproduce the J/$\psi$ $R_{\rm AA}$ centrality dependence within uncertainties but underestimates the $\psi(2S)$ $R_{\rm AA}$ in the most central collisions. It is worth pointing out that both models are for prompt charmonia only, whereas the data are for inclusive charmonium measurements. In addition, both models use different input parameters, in particular, a different value of the charm production cross section, which makes the comparison to the data and between models not straightforward.

\begin{figure}[!htb]
\begin{center}
\vspace{9pt}
\includegraphics[scale=0.4]{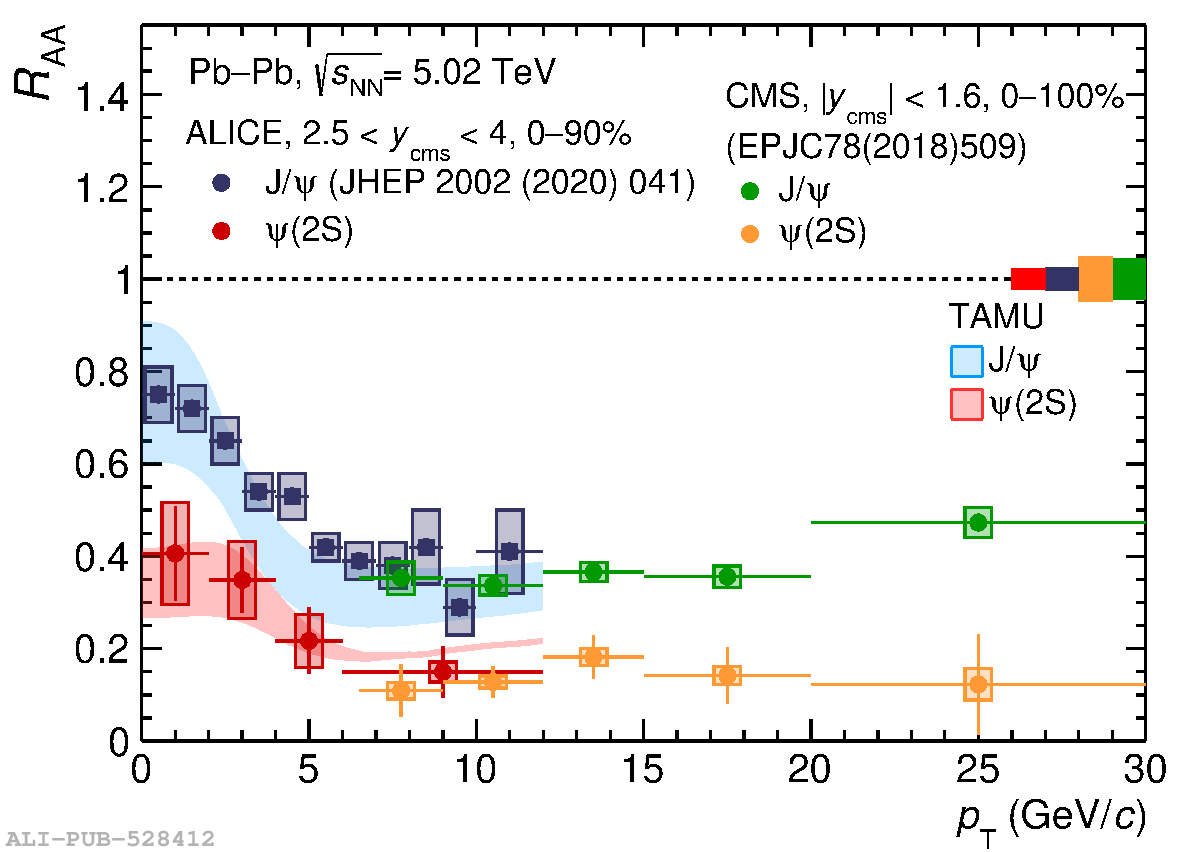}
\caption{Nuclear modification factor of the $\psi(2S)$~\cite{ALICE:2022jeh} and  J/$\psi$~\cite{ALICE:2016flj} as a function of $p_{\rm T}$. The values are compared with model predictions from TAMU~\cite{Du:2015wha} and with CMS measurements~\cite{CMS:2017uuv}.}
\label{fig:Figure2}
\end{center}
\end{figure}

The $p_{\rm T}$ dependence of the $\psi(2)$ $R_{\rm AA}$ is shown in Figure~\ref{fig:Figure2}, together with the J/$\psi$ $R_{\rm AA}$ and Transport Model predictions~\cite{Du:2015wha}. It is also compared with a similar measurement performed by CMS~\cite{CMS:2017uuv} in the $p_{\rm T}$ range $6 < p_{\rm T} < 30$~GeV/$c$, in the rapidity range $|y|<1.6$, and in the centrality range 0-100\%. Similarly to the J/$\psi$ case, the $\psi(2S)$ $R_{\rm AA}$ increases at low $p_{\rm T}$. In the common transverse momentum range, ALICE and CMS results are in very good agreement despite the different rapidity coverage, which allows one to see that the strong $\psi(2S)$ suppression persists up to $p_{T} = 30$~GeV/$c$. The transport model predictions can again reproduce the $p_{\rm T}$ dependence of both the J/$\psi$ and $\psi(2S)$ within uncertainties from 0 to 12~GeV/$c$. This strengthens the recombination scenario hypothesis, which dominates at low $p_{\rm T}$.

\begin{figure}[!htb]
\begin{center}
\vspace{9pt}
\includegraphics[scale=0.45]{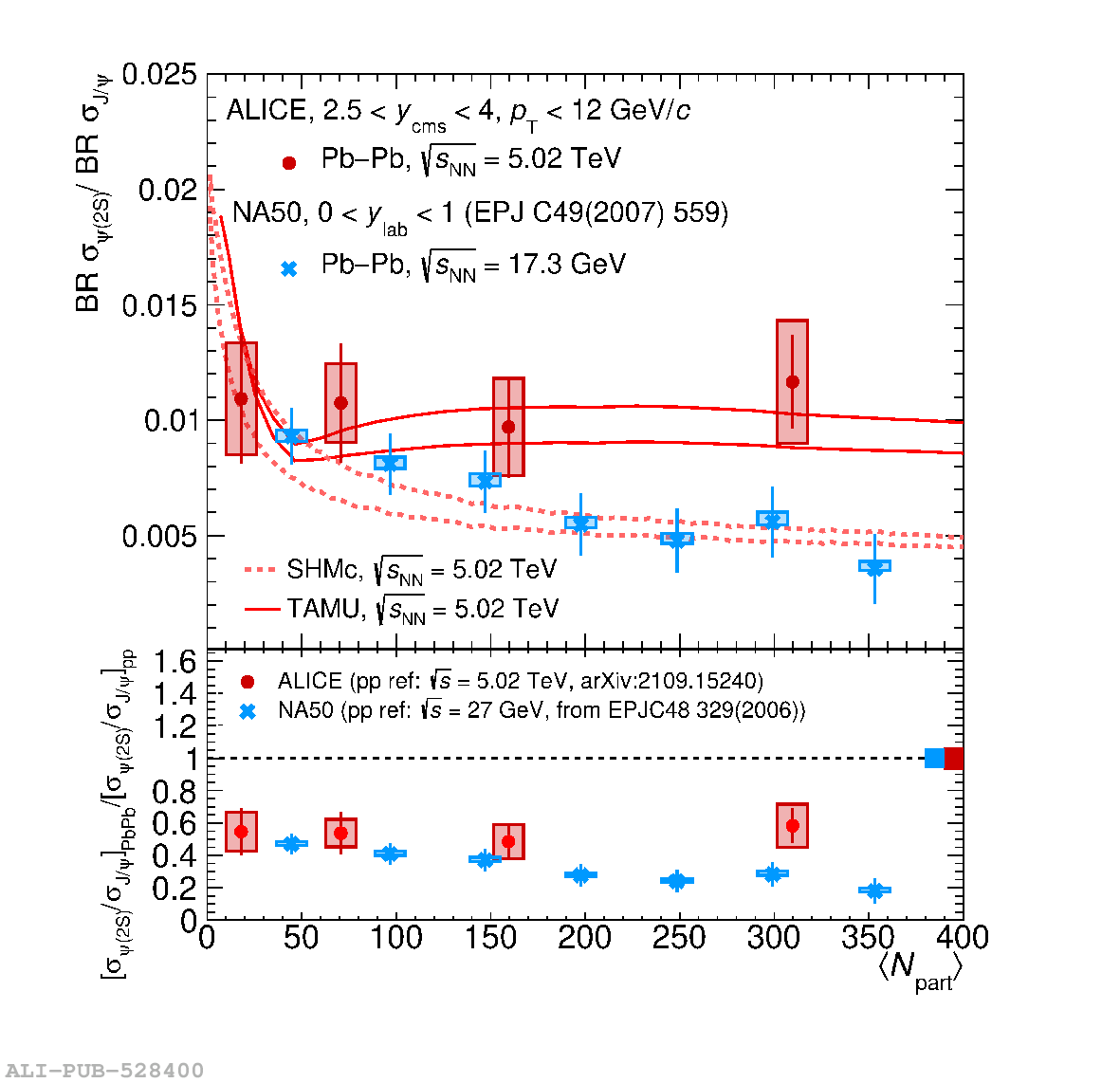}
\caption{Ratio of the $\psi(2S)$ and J/$\psi$ cross sections multiplied with the dimuon branching ratios as a function of centrality. In the lower panel the double ratio of the Pb--Pb and pp values is shown. The results are compared to predictions from the TAMU~\cite{Du:2015wha} and SHMc~\cite{Andronic:2019wva} models and to results of the SPS NA50 experiment~\cite{NA50:2006rdp}.}
\label{fig:Figure3}
\end{center}
\end{figure}

The $\psi(2S)$-to-J/$\psi$ yield ratios in Pb--Pb collisions are presented in Figure~\ref{fig:Figure3}. Using a ratio has the advantage that several systematic uncertainties being common to J/$\psi$ and $\psi(2S)$ cancel out, as well as common theoretical inputs to/$\psi$ and $\psi(2S)$ calculations in models. It is compared with SHMc~\cite{Andronic:2019wva} and TAMU~\cite{Du:2015wha} predictions and with the NA50~\cite{NA50:2006rdp} measurements at $\sqrt{s_{\rm NN}} = 17.3$~GeV, a much lower collision energy.

The $\psi(2S)$-to-J/$\psi$ yield ratio shows no centrality dependence. The TAMU model prediction shows good agreement with the data in the entire centrality range. The SHMc model can reproduce the data points for the peripheral and semi-central collisions but underestimates the ratio in the most central Pb--Pb collisions by about $1.86\sigma$. In the bottom panel of Figure~\ref{fig:Figure3}, the double ratio, i.e. the cross-section ratio in Pb--Pb collisions normalized by the cross-section ratio in pp collisions, is compared with the NA50 measurement~\cite{NA50:2006rdp} of the double ratio. The NA50 measurement is lower in most central collisions for both the single and double ratio than the ALICE measurement. The double ratio shows that the $\psi(2S)$ suppression in Pb--Pb collisions with respect to pp collisions is larger than the J/$\psi$ one by a factor of 2.

\begin{figure}[!htb]
\begin{center}
\vspace{9pt}
\includegraphics[scale=0.45]{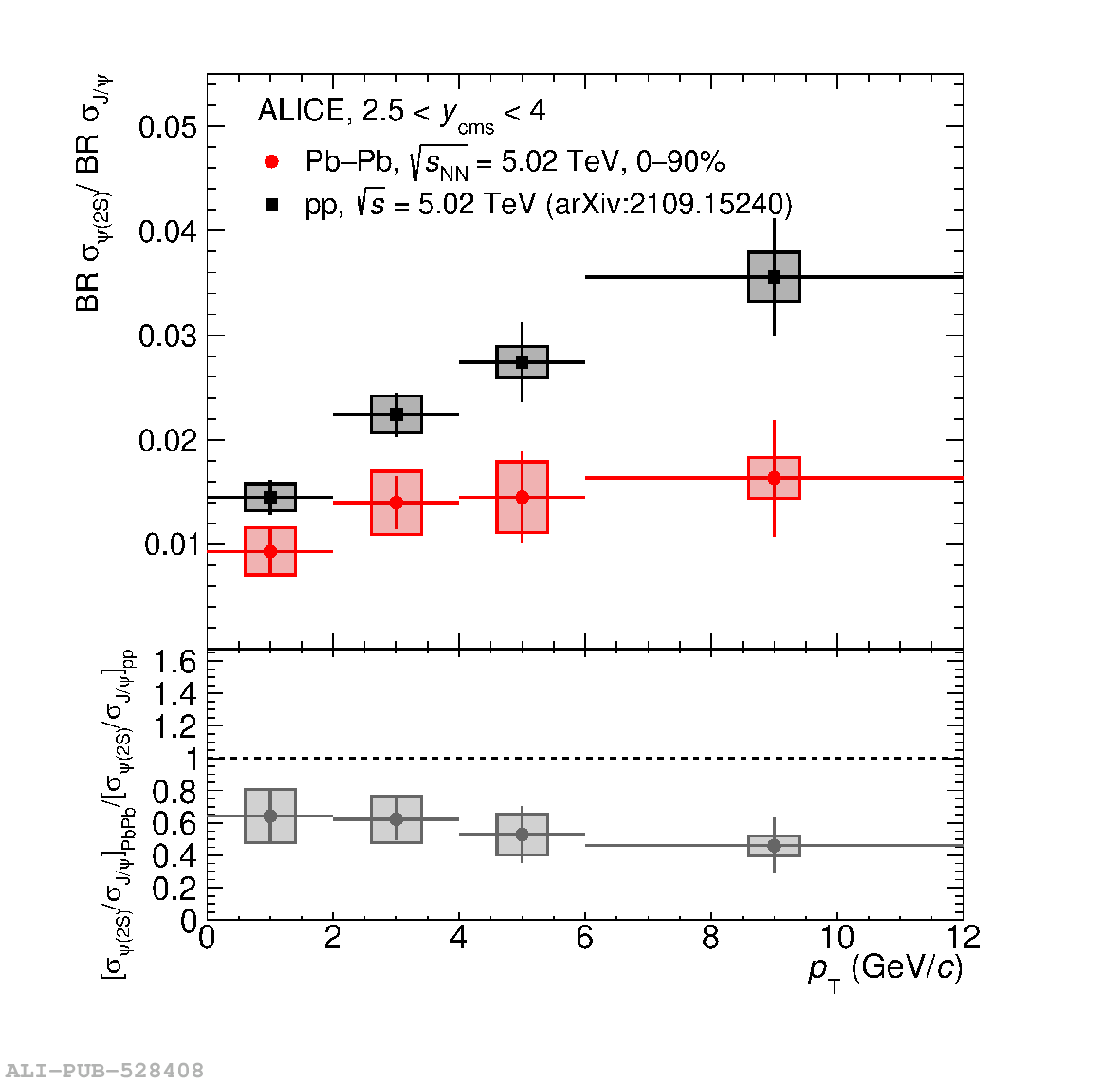}
\caption{Ratio of the $\psi(2S)$ and J/$\psi$ cross sections multiplied with the dimuon branching ratios as a function of $p_{\rm T}$, in Pb--Pb~\cite{ALICE:2022jeh} and pp collisions~\cite{ALICE:2021qlw}. In the lower panel the double ratio of the Pb--Pb and pp values is shown.}
\label{fig:Figure4}
\end{center}
\end{figure}

Finally, the single and double ratios as a function of $p_{\rm T}$ are shown in Figure~\ref{fig:Figure4}. The  $\psi(2S)$-to-J/$\psi$ yields ratio in Pb--Pb is compared with the ratio in pp collisions. Both ratios increase as a function of $p_{\rm T}$, but the Pb--Pb ratio tends to show a smaller rise as a function of $p_{\rm T}$ than the pp one, although, not very significant given the experimental uncertainties. . The double ratio values (bottom plot) might hint at a decrease with $p_{\rm T}$, down to a value of around 0.5, indicating a possible increase in the relative suppression of the $\psi(2S)$.

In summary, the inclusive nuclear modification factor of the $\psi(2S)$ in Pb--Pb collisions at $\sqrt{s_{\rm NN}} = 5.02$~TeV and forward rapidity has been measured down to $p_{\rm T} = 0$. The results show a clear suppression hierarchy, with the $\psi(2S)$ being more suppressed than the J/$\psi$ in the entire centrality and $p_{\rm T}$ range. The $\psi(2S)$ shows an increase of the $R_{\rm AA}$ at low $p_{\rm T}$, which is compatible with theoretical predictions that include a recombination scenario. SHMc and TAMU models are both able to reproduce the centrality dependence of the $R_{\rm AA}$ fairly well, even if the TAMU model shows a better agreement with the data in the most central collisions. These measurements will be improved in the coming years with the data from the ongoing Run3 at the LHC, and the addition of the Muon Forward Tracker that allows one to distinguish between the prompt and non-prompt charmonium production at forward rapidity.

{\setstretch{0.5}
\bibliographystyle{JHEP}
\bibliography{my-bib-database}
}

\end{document}